\documentclass{iopart}
\usepackage{graphicx}  
\usepackage{latexsym}
\usepackage{amssymb}

\jl{6}        
\eqnobysec    

\def\beq{\begin{equation}}
\def\eeq{\end{equation}}
\def\rmd{{\rm d}} 
\def\rmD{{\rm D}}

\def\version{\today}

\begin{document}

\begin{flushright}
Current version: \version 
\end{flushright}

\title[Spinning bodies and the Poynting-Robertson effect]
{Spinning bodies and the Poynting-Robertson effect in the Schwarzschild spacetime}

\author{
Donato Bini$^* {}^\S{}^\dag$ and  
Andrea Geralico${}^\S{}^\ddag$
}
\address{
  ${}^*$\
Istituto per le Applicazioni del Calcolo ``M. Picone,'' CNR, I--00185 Rome, Italy
}
\address{
  ${}^\S$\
  ICRA,
  University of Rome ``La Sapienza,'' I--00185 Rome, Italy
}
\address{
  ${}^\dag$\
  INFN sezione di Firenze, I--00185
  Sesto Fiorentino (FI), Italy
}
\address{
  $^\ddag$
  Physics Department,
  University of Rome ``La Sapienza,'' I--00185 Rome, Italy
}

\begin{abstract}
A spinning particle in the Schwarzschild spacetime deviates from geodesic behavior 
because of its spin. A spinless particle also deviates from geodesic behavior when a test radiation field  is superimposed on the Schwarzschild background: in fact the interaction with the radiation field, i.e., the absorption and re-emission of radiation, leads to a friction-like drag force 
responsible for the well known effect which exists already in Newtonian gravity, the Poynting-Robertson effect.
Here the Poynting-Robertson effect is extended to the case of spinning particles
by modifying the Mathisson-Papapetrou model describing the motion of spinning test particles to  account for the contribution of the radiation force. The resulting equations are numerically integrated and some typical orbits are shown in comparison with the spinless case.
Furthermore, the interplay between spin and radiation forces is discussed by analyzing the deviation from circular geodesic motion on the equatorial plane when the contribution due to the radiation can also be treated as a small perturbation.
Finally the estimate of the amount of radial variation from the geodesic radius is  shown to be measurable in principle. 
\end{abstract}

\pacno{04.20.Cv}

\section{Introduction}

The motion of classical spinning test particles in a given gravitational background is described by the well known
Mathisson-Papapetrou (MP) model \cite{math37,papa51}.
Let $U^\alpha=\rmd x^\alpha /\rmd\tau$ be the timelike unit tangent vector to  the ``center of mass line'' ${\mathcal C}_U$ of the spinning particle  used to perform the multipole reduction, parametrized by the proper time $\tau$. 
The equations of motion are
\begin{eqnarray}
\label{papcoreqs1}
\frac{\rmD P^{\mu}}{\rmd \tau}&=&-\frac12R^{\mu}{}_{\nu\alpha\beta}U^{\nu}S^{\alpha\beta}\equiv F^{\rm (spin)}{}^{\mu}\,, \\
\label{papcoreqs2}
\frac{\rmD S^{\mu\nu}}{\rmd \tau}&=&P^{\mu}U^{\nu}-P^{\nu}U^{\mu}\,,
\end{eqnarray}
where $P^{\mu}$ is the total 4-momentum of the particle and the antisymmetric tensor $S^{\mu\nu}$ denotes the spin tensor (intrinsic angular momentum) associated with it; both fields are defined only along this center of mass world line. 
This system of 10 equations evolves $P$ and $S$ along ${\mathcal C}_U$ but contains 13 unknown quantities: $U$ (3), $P$ (4), $S$ (6).
Consistency of the model is ensured by imposing the Tulczyjew-Dixon \cite{tulc59,dixon64,dixon69,dixon70,dixon73,dixon74} supplementary conditions  
\beq
\label{Tconds}
S^{\mu\nu}P_\nu=0\,.
\eeq
Moreover, implicit in the model is the requirement that the spin structure of the particle should produce very small deviations from geodesic motion, in the sense that the length scale naturally associated with the spin should be very small when compared with the one associated with the curvature tensor of the spacetime itself. 
Otherwise, large values of spin would require taking into account the particle backreaction on the spacetime metric, i.e., the problem should be approached from a completely different point of view.

Let us consider a spinless test particle orbiting a star which emits radiation.
The radiation pressure of the light emitted by the star, in addition to the direct effect of the outward radial force,
exerts a drag force on the particle's motion.  This usually causes the body to fall into the star unless it is so small that the radiation pressure pushes it away from the star, a phenomenon called the Poynting-Robertson effect since it was first investigated by Poynting \cite{poynting} using Newtonian gravity and then calculated in the framework of linearized general relativity by Robertson \cite{robertson}. Successively many authors studied the Poynting-Robertson effect in more concrete situations, starting from the case of slowly evolving elliptical orbits for meteors \cite{wyatt}, to more recent works \cite{gues, abr-ell-lan,lam-mil,mil-lam3,mil-lam1}, where rotation of the emitting star is taken into account. 
More recently the Poynting-Robertson effect for a spinless test particle orbiting a black hole was studied in both the Schwarzschild and Kerr spacetimes \cite{bijanste} without the restriction of slow motion, but ignoring the finite size of the radiating body. 

Here we generalize the above discussion to the more realistic case of a spinning test particle subject to the  Poynting-Robertson effect by including the radiation forces in the Mathisson-Papapetrou model.
Today we have evidence of the existence of accreting matter around massive compact objects, e.g., active galactic nuclei \cite{agn}. The dynamical behavior of particles in close orbits around massive objects, while interacting with their radiation field, can be relevant when studying the evolution of shell or disk-like configurations of dust around intense radiative relativistic sources, where the loss of angular momentum via the Poynting-Robertson effect could act as a dust accretion mechanism. 
The radiation mechanism around a real accreting compact object is generally very complicated. 
We will limit ourselves to the case of a coherent flux of photons traveling along geodesics in some preferred direction.
Possible scenarios include a hot neutron star, a black hole accreting radiation or a system with an accretion disk which radiates.  
When the approximation of point-like test particles is no longer valid, we expect that the real Poynting-Robertson effect offers a different behavior with respect to the standard Poynting-Robertson effect for small dust particles.
The Mathisson-Papapetrou model allows us to take into account the actual size of the particle in the framework of general relativity by introducing a characteristic length of the particle itself through its spin.

It is worth mentioning that there also exists a wide literature concerning pseudo-classical test spinning particles, whose equations of motion reduce under certain limit to the classical MP equations \cite{grassberger,berezin,barducci,vanholten}. 
In fact, spinning particles can be equivalently described by pseudo-classical mechanics models in which the spin degrees of freedom are characterized in terms of anticommuting Grassmann variables, associated --- in the semiclassical limit --- with the components of the spin tensor of the particle.  
The Lagrangian formulation can be used as well to study spinning particle motion in external fields.
Here, however, we consider only classical test bodies, leaving for future work further generalizations of this analysis to the case of pseudo-classical particles.

\section{Motion in the Schwarzschild spacetime}

Consider a Schwarzschild spacetime, whose line element written in standard coordinates is given by
\beq 
\label{metric}
\rmd  s^2 = -N^2\rmd t^2 + N^{-2} \rmd r^2 
+ r^2 (\rmd \theta^2 +\sin^2 \theta \rmd \phi^2)\,,
\eeq
where $N=(1-2M/r)^{1/2}$ denotes the lapse function, and introduce the usual orthonormal frame adapted to the static observers (or Zero Angular Momentum Observers, ZAMOs) following the time lines
\beq
\label{frame}
e_{\hat t}=N^{-1}\partial_t\,, \quad
e_{\hat r}=N\partial_r\,, \quad
e_{\hat \theta}=\frac{1}{r}\partial_\theta\,, \quad
e_{\hat \phi}=\frac{1}{r\sin \theta}\partial_\phi\,,
\eeq
where $\{\partial_t, \partial_r, \partial_\theta, \partial_\phi\}$ is the coordinate frame.

We limit our analysis to  the equatorial plane $\theta=\pi/2$, where the situation is relatively receptive to analytical treatment.
As a convention, the physical (orthonormal) 
component along $-\partial_\theta$ which is perpendicular to the equatorial plane will be referred to as ``along the positive $z$-axis" and will be indicated by the index $\hat z$, when convenient: $e_{\hat z}=-e_{\hat \theta}$.

\subsection{Test particles subject to PR effect}

Let a pure electromagnetic radiation field be superposed as a test field on the gravitational background described by the metric (\ref{metric}),  with the energy-momentum tensor
\beq
\label{ten_imp}
T^{\alpha\beta}=\Phi^2 k^\alpha k^\beta, \qquad k^\alpha k_\alpha=0\ ,
\eeq
where $k$ is assumed to be tangent to an  affinely parametrized outgoing null geodesic in the equatorial plane, i.e., $k^\alpha \nabla_\alpha k^\beta=0$ with $k^\theta=0$.
 
We will only consider photons in the equatorial plane which are
in outward radial motion with respect to the ZAMOs, namely with 4-momentum
\beq
\label{eq:phot}
k= E(n)(n+e_{\hat r})\,, 
\eeq
where $n=e_{\hat t}$ is the ZAMO $4$-velocity and $E(n)= E/N$ is the relative energy of the photon as seen by the ZAMOs. Here $E=-k_t$ is the conserved energy associated with the timelike Killing vector field $\partial_t$. Note also that $L=k_\phi=0$, i.e., the conserved angular momentum associated with the rotational Killing vector field $\partial_\phi$ is assumed to vanish. This is consistent with having a non-rotating light source in a non-rotating spacetime.

Since $k$ is completely determined, the coordinate dependence of the quantity $\Phi$  then follows 
from the conservation equations  $T^{\alpha\beta}{}_{;\beta}=0$, and will only depend on $r$ in the equatorial plane due to the axial symmetry.
We find
\beq
\Phi=\frac{\Phi_0}{r}\,.
\eeq

Consider now a test particle moving in the equatorial plane $\theta=\pi/2$ accelerated by the radiation field, i.e. with 4-velocity 
\beq\label{polarnu}
\fl\qquad 
U=\gamma(U,n) [n+ \nu(U,n)]\,,\quad 
\nu(U,n)\equiv \nu^{\hat r}e_{\hat r}+\nu^{\hat \phi}e_{\hat \phi} 
=  \nu (\sin \alpha e_{\hat r}+  \cos \alpha e_{\hat \phi})  \,,
\eeq
where $\gamma(U,n)=1/\sqrt{1-||\nu(U,n)||^2}\equiv\gamma$ is the Lorentz factor and the abbreviated notation $\nu^{\hat a}\equiv\nu(U,n)^{\hat a}$ has been used.  Similarly $\nu \equiv ||\nu(U,n)||$ and $\alpha$ are the magnitude of the spatial velocity $\nu(U,n)$ and its
polar angle  measured clockwise from the positive $\phi$ direction in the $r$-$\phi$ tangent plane respectively, while $\hat\nu\equiv\hat \nu(U,n)$ is the associated unit vector.
Note that $\alpha=0$ corresponds to azimuthal motion with respect to the ZAMOs, while 
$\alpha=\pm \pi/2$ corresponds to (outward/inward) radial motion with respect to the ZAMOs.

A straightforward calculation gives the coordinate components of $U$
\beq\fl\qquad
\label{Ucoord_comp}
U^t\equiv \frac{\rmd t}{\rmd \tau}=\frac{\gamma}{N}\,, \qquad 
U^r\equiv \frac{\rmd r}{\rmd \tau}=\gamma N\nu^{\hat r}\,, \qquad 
U^\phi\equiv \frac{\rmd \phi}{\rmd \tau}=\frac{\gamma \nu^{\hat \phi}}{r}\,,
\eeq
where $\tau$ is the proper time parameter along ${\mathcal C}_U$, and $U^\theta\equiv \rmd \theta/\rmd \tau=0$. 
Solving these for the magnitude and polar angle leads to
\begin{eqnarray}
\tan \alpha &=& 
\frac{1}{Nr}
\frac{\rmd r}{\rmd \phi}\,,\qquad 
\nu= 
\frac{1}{N^2}\sqrt{\left( \frac{\rmd r}{\rmd t} \right)^2
         + N^2r^2\left( \frac{\rmd \phi}{\rmd t} \right)^2} \,.
\end{eqnarray}

The scattering of radiation as well as the momentum-transfer cross section $\sigma$ (assumed to be a constant) of the particle are assumed to be independent of the direction and frequency of the radiation so that the associated force is given by  \cite{poynting,robertson,abr-ell-lan}
\beq
\label{frad}
F^{\rm (rad)}{}^\alpha = -\sigma {\mathcal P}(U)^\alpha{}_\beta \, T^{\beta}{}_\mu \, U^\mu\,,
\eeq
where ${\mathcal P}(U)^\alpha{}_\beta=\delta^\alpha_\beta+U^\alpha U_\beta$ projects orthogonally to $U$.
Explicitly 
\beq
F^{\rm (rad)}{}^\alpha=-\sigma \Phi^2 ({\mathcal P}(U)^\alpha{}_\beta k^\beta)\, (k_\mu U^\mu)\,,
\eeq
implying
\beq\fl
F^{\rm (rad)}=\frac{mA}{N^2r^2}\gamma^3 (1-\nu^{\hat r})\left[(\nu^{\hat r}-\nu^2)n
+\left(1-\nu^{\hat r} -(\nu^{\hat \phi})^2\right)e_{\hat r}
-(1-\nu^{\hat r})\nu^{\hat \phi}e_{\hat \phi}\right]\,,
\eeq
where we have used the notation $\sigma \Phi_0^2 E^2=mA$, as in Ref. \cite{bijanste}.
Test particle motion is then described by the equation
\beq
\label{eqPRtest}
m a(U)^\mu\equiv m \frac{\rmD U^\mu}{\rmd\tau}=F^{\rm (rad)}{}^\mu\,,
\eeq
and has been studied in detail in Ref. \cite{bijanste}.

\subsection{Generalization to spinning particles}

The most direct and simple generalization of Eq.~(\ref{eqPRtest}) to the case of spinning test particles consists in including the radiation force term in Eq.~(\ref{papcoreqs1}), so that one has
\beq
\label{PRspin}
\frac{\rmD P^{\mu}}{\rmd \tau}=F^{\rm (spin)}{}^{\mu}+F^{\rm (rad)}{}^\mu\,, \\
\eeq
plus the additional relations (\ref{papcoreqs2}) and (\ref{Tconds}) involving the spin.
Let us proceed to analyse the motion  of spinning particles subject to the Poynting-Robertson effect in the equatorial plane of Schwarzschild spacetime.

The 4-momentum $P=m u$ for motion in the equatorial plane is
\beq
\label{Updef}
u=\gamma_u[n+\nu_u(\sin\alpha_u e_{\hat r}+\cos\alpha_u e_{\hat \phi})]\,, \qquad
\gamma_{u}=\frac1{\sqrt{1-\nu_{u}^2}}\,,
\eeq
and introduce the spin vector associated with $S_{\mu\nu}$ by spatial duality
\beq
S^\beta=u_\alpha \eta^{\alpha\beta\mu\nu}S_{\mu\nu}\,,
\eeq 
where $\eta_{\alpha\beta\gamma\delta}=\sqrt{-g} \epsilon_{\alpha\beta\gamma\delta}$ is the unit volume 4-form and $\epsilon_{\alpha\beta\gamma\delta}$ ($\epsilon_{0123}=1$) is the Levi-Civita alternating symbol. 
It is also useful to consider the scalar invariant
\beq
\label{sinv}
s^2=\frac12 S_{\mu\nu}S^{\mu\nu}\,, 
\eeq
constant along ${\mathcal C}_U$ because of Eqs.~(\ref{papcoreqs2}) and (\ref{Tconds}).
Consistency of the model requires that the length scale $|s|/m$ associated with the spinning particle be much smaller than the one associated with the background spacetime, say $M$, namely
\beq
|\hat s|\equiv \frac{|s|}{mM}\ll 1\,.
\eeq

Let us consider Eqs.~(\ref{papcoreqs1}) and (\ref{papcoreqs2}) with $U^{\alpha}$ given by Eq.~(\ref{Ucoord_comp}) and $u^{\alpha}$ given by Eq.~(\ref{Updef}).
In the spinless case $P$ is aligned with $U$, i.e., $u=U$, implying that $\nu=\nu_u$.
The presence of the spin causes a change in both $U$ and $u$ according to
\beq
U=U_0+{\hat s}U_{\hat s}\,, \qquad
u=U_0+{\hat s}u_{\hat s}\,, 
\eeq
where 
\beq
U_0=\gamma_0(n+\nu_0^{\hat r}e_{\hat r}+\nu_0^{\hat \phi}e_{\hat \phi})\
\eeq
satisfies Eq.~(\ref{eqPRtest}) and corrections are first order in the spin.
Higher order terms in Eqs.~(\ref{papcoreqs1}) and (\ref{papcoreqs2}) are neglected.
This leads to two different sets of equations for zeroth and first order in spin respectively, which are listed in \ref{app1}.

The spin force to first order in $\hat s$ is given by
\beq
F^{\rm (spin)}=-\frac{3mM^2}{r_0^3}{\hat s}\gamma_{0}^2\nu^{\hat \phi}_0(\nu^{\hat r}_0n+e_{\hat r})\,.
\eeq
We find that the mass of the spinning particle $m$ is a constant of motion.
Furthermore, from the evolution equations for the spin it follows that the spin vector has a single nonvanishing and constant component along $\theta$ (or $z$), namely
\beq
S=-S^{\hat\theta}e_{\hat \theta}=se_{\hat z}\,. 
\eeq

\begin{figure}
\begin{center}
\includegraphics[scale=0.5]{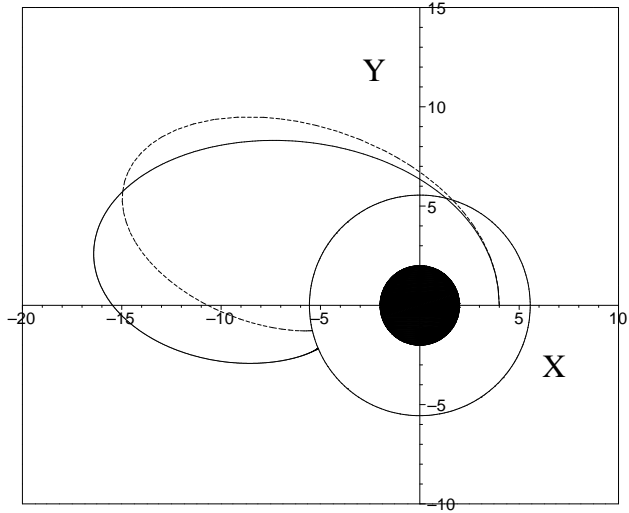}
\end{center}
\caption{The orbit of a spinning particle (solid curve) subject to the Poynting-Robertson effect is shown for the choice of parameters $A/M=0.8$ and $\hat s=0.5$ ($X=r\cos\phi$ and $Y=r\sin\phi$ are Cartesian-like coordinates).
The starting point is located at $r_0(0)=4M$ and $\phi_0(0)=0$ with $\nu_{u0}(0)=0.7$, $\alpha_{u0}(0)=0$, $t_s(0)=0$, $r_s(0)=0$ and $\phi_s(0)=0$, $\nu^{\hat r}_s(0)=0$ and $\nu^{\hat \phi}_s(0)=0$.
The values of the spin parameter has been exaggerated in order to distinguish the difference from the motion of a spinless particle (dashed curve).
The inner circle is the horizon $r=2M$, while the outer circle is at the critical radius $r_{\rm(crit)}=5.5M$ which is outside the initial data position.
}
\label{fig:1}
\end{figure}

\begin{figure}
\begin{center}
\includegraphics[scale=0.5]{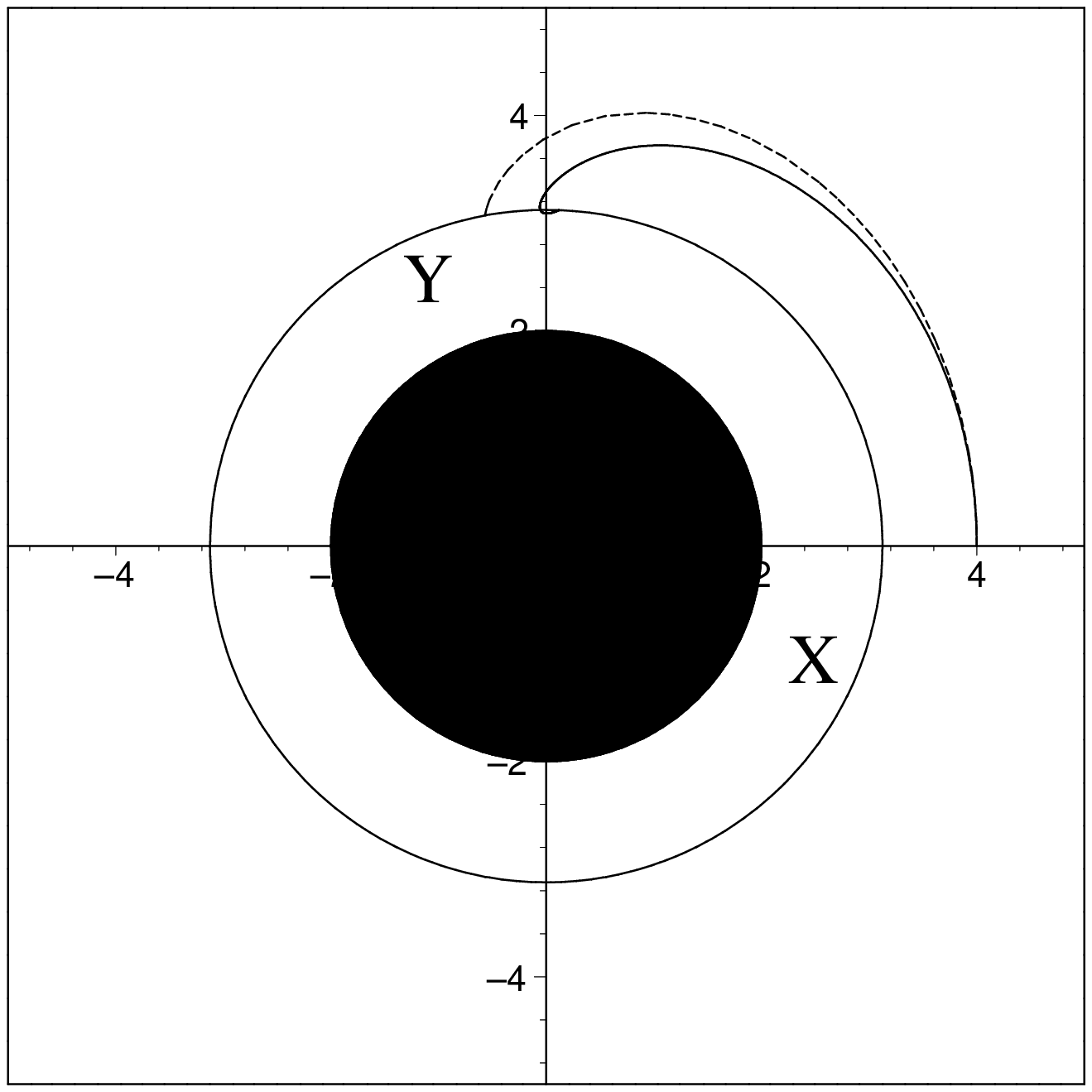}
\end{center}
\caption{The orbit of a spinning particle (solid curve) subject to the Poynting-Robertson effect is shown for the choice of parameters $A/M=0.6$ and $\hat s=0.5$.
The starting point is located at $r_0(0)=4M$ and $\phi_0(0)=0$ with $\nu_{u0}(0)=0.5$, $\alpha_{u0}(0)=0$, $t_s(0)=0$, $r_s(0)=0$ and $\phi_s(0)=0$, $\nu^{\hat r}_s(0)=0$ and $\nu^{\hat \phi}_s(0)=0$.
The corresponding orbit for a spinless particle is also shown (dashed curve).
The critical radius $r_{\rm(crit)}=3.125M$ is inside the initial data position. 
}
\label{fig:2}
\end{figure}

\begin{figure}
\begin{center}
\includegraphics[scale=0.5]{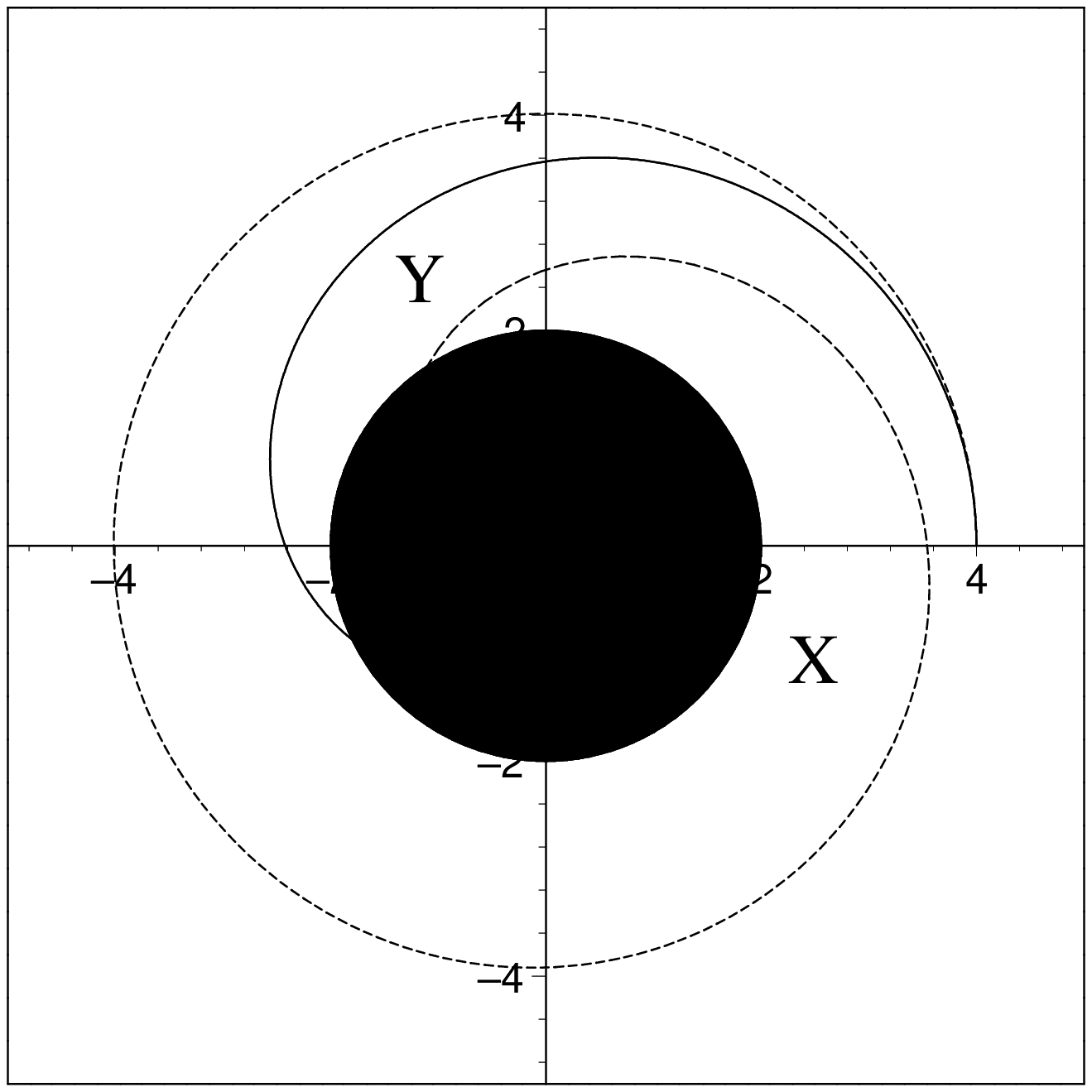}
\end{center}
\caption{The orbit of a spinning particle (solid curve) subject to the Poynting-Robertson effect is shown for the choice of parameters $A/M=0.01$ and $\hat s=0.5$.
The starting point is located at $r_0(0)=4M$ and $\phi_0(0)=0$ with $\nu_{u0}(0)=\nu_K\approx0.7071$, $\alpha_{u0}(0)=0$, $t_s(0)=0$, $r_s(0)=0$ and $\phi_s(0)=0$, $\nu^{\hat r}_s(0)=0$ and $\nu^{\hat \phi}_s(0)=0$.
The corresponding orbit for a spinless particle is also shown (dashed curve).
In this case $r_{\rm(crit)}\approx 2M$.
}
\label{fig:3}
\end{figure}

Figures \ref{fig:1}--\ref{fig:3} show some numerical solutions for the orbits in the strong field region. 
Of course it only makes sense to consider the exterior solutions for radii larger than some minimum radius $R$ outside the horizon in order to model the geometry outside a star (or some other physical source) of radius $R$ producing the outflow of radiation. 

In the case of spinless particles there exists a condition representing the balancing of the gravitational attraction and the radiation pressure at constant $r_0$ and $\phi_0$, namely
\beq
\label{balcond}
\frac{A}{M}= \left(1-\frac{2M}{r_0}\right)^{1/2} 
\quad\rightarrow\quad
r_0= r_{\rm(crit)} \equiv  \frac{2M}{1-A^2/M^2}
\,.
\eeq
This behavior also characterizes the motion of spinning particles as well, as shown in \ref{app1a}.
Figs.~\ref{fig:1} and \ref{fig:2} show some typical solution curves starting initially with purely azimuthal velocity either inside (Fig.~\ref{fig:1}) or outside (Fig.~\ref{fig:2}) the critical radius at which a particle initially at rest with respect to the ZAMOs (which in turn are at rest with respect to the coordinate system) remains at rest. For comparison, the corresponding curves for spinless particles with the same initial data are also shown.
If $A/M\ll1$, the critical radius approaches the horizon $r_{\rm(crit)}\approx 2M$. For instance, an initially Keplerian circular orbit gradually spirals towards the central source, as illustrated in Fig.~\ref{fig:3}.


\section{Deviation from the circular geodesic}

Consider now the corrections to geodesic circular motion, by taking the effect of the radiation field to also be small.

In the absence of both spin and radiation we assume the geodesic motion of the particle to be circular at $r=r_0$ ($r_0>3M$ in order $U_K$ to be timelike), that is
\beq
\label{Ugeo}
U=U_K=\gamma_K (n \pm \nu_K e_{\hat \phi})\,,
\eeq
where the Keplerian value of speed ($\nu_K$) and the associated Lorentz factor ($\gamma_K$) and angular velocity ($\zeta_K$) are given by
\beq\fl\qquad
\label{nuKdef}
\nu_K=\sqrt{\frac{M}{r_0-2M}}\,,\qquad 
\gamma_K=\sqrt{\frac{r_0-2M}{r_0-3M}}\,,\qquad 
\zeta_K=\sqrt{\frac{M}{r_0^3}}\,.
\eeq
The $\pm$ signs in Eq.~(\ref{Ugeo}) correspond to co-rotating $(+)$ or counter-rotating $(-)$ orbits with respect to increasing values of the azimuthal coordinate $\phi$ (counter-clockwise motion as seen from above). 
The azimuthal direction in the local rest space of $U_K$ pointing in the direction of relative  motion (i.e., the boost of $e_{\hat \phi}$ in the local rest space of $U_K$) is specified by the following unit vector orthogonal to $U_K$ in the $t$-$\phi$ plane
\beq
{\bar U}_K=\gamma_K(\nu_Kn\pm e_{\hat \phi})\,,
\eeq
where the $\pm$ signs are correlated with those in $U_K$.

The parametric equations of $U_K$ are
\begin{eqnarray}
\label{circgeos}
t_K&=&t_0+\frac{\gamma_K}{N_0}\tau\equiv t_0+\Gamma_K \tau\,,\nonumber\\ 
{\phantom{x}}r&=&r_0\,,\qquad 
\theta=\frac{\pi}{2}\,,\nonumber\\
\phi_K&=&\phi_0\pm \frac{\gamma_K\nu_K}{r_0}\tau\equiv \phi_0\pm \Omega_K \tau\,,
\end{eqnarray}
where now $t_0$, $r_0$ and $\phi_0$ are constants and
\beq
\Gamma_K=\sqrt{\frac{r_0}{r_0-3M}}\,, \qquad
\Omega_K=\frac{1}{r_0}\sqrt{\frac{M}{r_0-3M}}\,.
\eeq

It is convenient to introduce a friction parameter $f$, so that the length scale $A$ associated with the radiation field is much smaller than $M$, i.e.,
\beq
f\equiv \frac{A}{M}\ll 1\,.
\eeq
Therefore, in the present analysis corrections to geodesic motion will be limited to first order terms in both parameters $\hat s$ and $f$, according to
\begin{eqnarray}\fl\qquad
\label{neargeo}
t&=&t_K+ft_f+{\hat s}t_{\hat s}\,, \quad
r=r_0+fr_f+{\hat s}r_{\hat s}\,, \quad
\phi=\phi_K+f\phi_f+{\hat s}\phi_{\hat s}\,, \nonumber\\
\fl\qquad
\nu^{\hat r}&=&f\nu^{\hat r}_f+{\hat s}\nu^{\hat r}_{\hat s}\,, \quad
\nu^{\hat \phi}=\pm\nu_K+f\nu^{\hat \phi}_f+{\hat s}\nu^{\hat \phi}_{\hat s}\,, \nonumber\\
\fl\qquad
\nu_u&=&\pm\nu_K+f\nu_{uf}+{\hat s}\nu_{u{\hat s}}\,, \quad
\alpha_u=f\alpha_{uf}+{\hat s}\alpha_{u{\hat s}}\,, 
\end{eqnarray}
where $t_K$ and $\phi_K$ are given by Eq.~(\ref{circgeos}).
This implies
\beq
\label{Uexp}
U=U_K+fU_f+{\hat s}U_{\hat s}\,,
\eeq
where
\begin{eqnarray}
U_f&=&\left(-\nu_K\frac{r_f}{r_0}\pm\gamma_K^2\nu_f^{\hat \phi}\right){\bar U}_K+\gamma_K\nu_f^{\hat r}e_{\hat r}\,,\nonumber\\
U_{\hat s}&=&\left(-\nu_K\frac{r_{\hat s}}{r_0}\pm\gamma_K^2\nu_{\hat s}^{\hat \phi}\right){\bar U}_K+\gamma_K\nu_{\hat s}^{\hat r}e_{\hat r}\,.
\end{eqnarray}
Similarly
\beq
u=U+fu_f+{\hat s}u_{\hat s}\,,
\eeq
with $U$ given by Eq.~(\ref{Uexp}) and
\beq
u_f=\gamma_K(\nu_K^2\nu_f^{\hat \phi}-\nu_f^{\hat r})e_{\hat r}\,,\qquad
u_{\hat s}=\gamma_K(\nu_K^2\nu_{\hat s}^{\hat \phi}-\nu_{\hat s}^{\hat r})e_{\hat r}\,,
\eeq
as discussed in \ref{app2}.

\begin{figure}
\begin{center}
\includegraphics[scale=0.45]{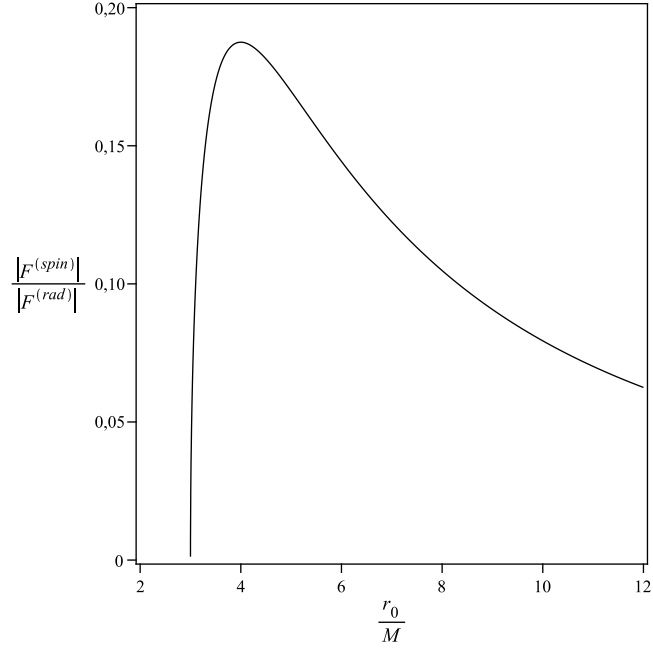}
\end{center}
\caption{The ratio between the magnitudes of spin and radiation forces given by Eq.~(\ref{ratio}) is plotted in units of $|{\hat s}|/f$ as a function of $r_0/M$.
}
\label{fig:4}
\end{figure}

To first order in $\hat s$ and $f$ the spin force and radiation force are given by
\begin{eqnarray}
F^{\rm (spin)}&=&\mp3mM{\hat s}\gamma_K^2 \zeta_K^2\nu_K e_{\hat r}\,,\nonumber \\
F^{\rm (rad)}&=& -mf\Omega_K\nu_K(\gamma_K\nu_K{\bar U}_K -e_{\hat r})\,,
\end{eqnarray}
respectively.
The ratio between the magnitudes of these forces has the form
\beq
\label{ratio}
\frac{|F^{\rm (spin)}|}{|F^{\rm (rad)}|}=3\frac{|\hat s|}{f}\left(\frac{M}{r_0}\right)^{3/2}\sqrt{1-\frac{3M}{r_0}}\,.
\eeq
Its behavior as a function of $r_0$ in units of $|{\hat s}|/f$ is shown in Fig.~\ref{fig:4}.


The equations governing first order perturbations are listed in \ref{app2}.
The corresponding solution is given by
\begin{eqnarray}\fl\quad
\label{solfs1}
\nu^{\hat r}_{\hat s}&=&\mp\frac{3M^2}{r_0^2}\frac{\Omega_K}{\Omega_{\rm ep}}\sin(\Omega_{\rm ep}\tau)\,, \nonumber\\
\fl\quad
\nu^{\hat r}_f&=&\frac{\nu_K^2}{r_0\Omega_{\rm ep}}\left\{\sin(\Omega_{\rm ep}\tau)+2r_0\zeta_K\frac{\Omega_K}{\Omega_{\rm ep}}[\cos(\Omega_{\rm ep}\tau)-1]\right\}\,, \nonumber\\
\fl\quad
\nu^{\hat \phi}_{\hat s}&=&-\frac{3M\zeta_K^3}{\Omega_{\rm ep}^2}\nu_K[\cos(\Omega_{\rm ep}\tau)-1]\,, \nonumber\\
\fl\quad
\nu^{\hat \phi}_f&=&\pm\frac{\nu_K^3}{r_0\Omega_{\rm ep}^2}\left\{\frac{\zeta_K}{r_0\Omega_{\rm ep}}[\cos(\Omega_{\rm ep}\tau)-1]-\frac{2\zeta_K^2}{\Omega_{\rm ep}}\sin(\Omega_{\rm ep}\tau)+\Omega_K^2\tau\right\}\,,
\end{eqnarray}
and
\begin{eqnarray}\fl\quad
\label{solfs2}
t_{\hat s}&=&\mp\frac{6M^2}{r_0}\frac{\Omega_K^3}{\Omega_{\rm ep}^3}[\sin(\Omega_{\rm ep}\tau)-\Omega_{\rm ep}\tau]\,, \nonumber\\
\fl\quad
t_f&=&4r_0\zeta_K\nu_K^2\frac{\Omega_K^3}{\Omega_{\rm ep}^4}\left\{[\cos(\Omega_{\rm ep}\tau)-1]+\frac{\Omega_{\rm ep}}{2r_0\zeta_K\Omega_K}[\sin(\Omega_{\rm ep}\tau)-\Omega_{\rm ep}\tau]+\frac38\gamma_K^2\Omega_{\rm ep}^2\tau^2\right\}\,, \nonumber\\
\fl\quad
r_{\hat s}&=&\pm3r_0\frac{\Omega_K\zeta_K}{\Omega_{\rm ep}^2}[\cos(\Omega_{\rm ep}\tau)-1]\,, 
\nonumber\\
\fl\quad
r_f&=&-r_0\zeta_K\frac{\Omega_K}{\Omega_{\rm ep}^2}\left\{[\cos(\Omega_{\rm ep}\tau)-1]-2r_0\zeta_K\frac{\Omega_K}{\Omega_{\rm ep}}[\sin(\Omega_{\rm ep}\tau)-\Omega_{\rm ep}\tau]\right\}\,, \nonumber\\
\fl\quad
\phi_{\hat s}&=&\pm\frac{\zeta_K}{\nu_K^2}t_{\hat s}\,, \qquad
\phi_f=\pm\frac{\zeta_K}{\nu_K^2}t_f\,,
\end{eqnarray}
where 
\beq
\label{omegaep}
\Omega_{\rm ep} = \sqrt{\frac{M (r_0-6M)}{r_0^3 (r_0 -3M)}}\
\eeq
is the well known epicyclic frequency governing the radial perturbations of circular geodesics.

The constant term in $r_{\hat s}$ represents the slight change in the radius of the circular orbit about which the solution oscillates with proper period $2\pi/\Omega_{\rm ep}$.
In contrast, the presence of a secular term in $r_f$ is responsible for the deviation from geodesic motion due to friction, which is measurable in principle.
In fact, by taking the mean values over a period of the perturbed radius we can estimate the amount of variation of the radial distance 
\beq\fl\qquad
\Big\langle\frac{\delta r}{r}\Big\rangle\equiv\frac{r-r_0}{r_0}=\Gamma_K\frac{\zeta_K^2}{\Omega_{\rm ep}^2}
\left[\left(1-2\pi r_0\zeta_K\frac{\Omega_K}{\Omega_{\rm ep}}\right)f\mp3M{\hat s}\gamma_K\zeta_KN_0\right]\,.
\eeq

For instance, for the motion of the Earth about the Sun we find 
\beq\fl\qquad
\Big\langle\frac{\delta r}{r}\Big\rangle\approx f\mp2\times10^{-17}\frac{(s/m)_\oplus}{\rm cm}
\approx3\times10^{-15}\mp4\times10^{-15}\approx10^{-15}\,,
\eeq
since $r_0\approx1.5\times10^{13}$ cm, $M=M_\odot\approx1.5\times10^{5}$ cm and the ratio $(s/m)_\oplus\approx200$ cm for the Earth;
the friction parameter is related to the ratio between the solar luminosity ${\mathcal L}_{\odot}\approx3.8\times10^{33}$ erg/s and the Eddington luminosity \cite{mil-lam1,bijanste} ${\mathcal L}_{\rm Edd}\approx1.3\times10^{38}$ erg/s, and for the Sun is given by $f\approx3\times10^{-15}$.
Therefore, in this case the effect of the radiation field on the orbit is of the same order as that due to spin.
Note that the estimate of the contribution due to spin is in agreement with \cite{mashsingh}.

The effect of the spin may become important when the orbiting extended body is a fast rotating object.
To illustrate the order of magnitude of the effect, we may consider the binary pulsar system PSR J0737-3039 as orbiting Sgr A$^*$, the supermassive ($M\simeq 10^6\ M_\odot$) black hole located at the Galactic Center \cite{falcke,muno}, at a distance of $r\simeq 10^9$ Km. 
The PSR J0737-3039 system consists of two close neutron stars (their separation is only $d_{AB} \sim 8 \times 10^5$ Km) of comparable masses $m_A\simeq 1.4\ M_\odot$, $m_B \simeq 1.2\ M_\odot$), but very different intrinsic spin period ($23$ ms of pulsar A vs $2.8$ s of pulsar B) \cite{lyne}. 
Its orbital period is about $2.4$ hours, the smallest yet known for such an object. Since the intrinsic rotations are negligible with respect to the orbital period, we can treat the binary system as a single object with reduced mass $\mu_{AB} \simeq 0.7 \ M_\odot$ and intrinsic rotation equal to the orbital period. 
The spin parameter thus turns out to be equal to $\hat s\approx1.0\times10^{-3}$.
The luminosity of Sgr A$^*$ is about $10^3{\mathcal L}_{\odot}$, whereas its Eddington luminosity is ${\mathcal L}_{\rm Edd}\approx10^{11}{\mathcal L}_{\odot}$, so that $f\approx10^{-18}$.
Therefore, in this case
\beq
\Big\langle\frac{\delta r}{r}\Big\rangle
\approx7.6\times10^{-19}\mp1.8\times10^{-7}\approx10^{-7}\,.
\eeq
Therefore, in this case the effect of the spin on the orbit dominates with respect to the friction due to the radiation field.

\section{Concluding remarks}

We have studied the motion of a classical spinning body in the field of a central radiating object. 
The model adopted is the standard Mathisson-Papapetrou model suitably modified by accounting for the contribution of the Poynting-Robertson radiation force in the equations of motion.
We have numerically integrated the whole set of Mathisson-Papapetrou equations in the case of equatorial motion and coherent flux composed of radially emitted photons.
The spin vector turns out to have only a constant nonvanishing component orthogonal to the motion plane.
We have shown some typical solution orbits in comparison with the spinless case.
The latter is characterized by the existence of a critical radius at which the balancing of the gravitational attraction and the radiation pressure occurs at constant radial and azimuthal coordinates depending on the strenght of the radiation field.
This feature has been proved to be maintained also in presence of spin. 
Dust particles would congregate at this radius leading to rings of matter to form.

Furthermore, we have discussed the interplay between spin and radiation forces by analyzing the deviation from circular geodesic motion on the equatorial plane when also the contribution due to friction can be treated as a small perturbation.
The features of the motion thus depend on two different parameters, the spin parameter and the friction parameter, which are taken as small in order to avoid backreaction. 
The presence of the spin causes a slight change in the radius of the circular orbit about which the solution oscillates and an increase/decrease of the angular velocity depending on whether the particle is co/counter rotating (i.e., moving clockwise or anticlockwise with respect to the positive $\phi$ direction, respectively).
In contrast, the presence of a secular term in the radial deviation due to friction determines a spiraling behavior of the orbit. 
This leads to a (average) radial variation from the geodesic radius whose amount is measurable, at least in principle.

The model presented here allows to account for the finite size of the particle subject to the Poynting-Robertson effect in a framework which is genuinely relativistic.
Obviously, in order to be physically more realistic, the model should be further generalized to take into account, for instance, the finite size of the radiating source and the contribution of higher order multipoles in the description of the actual size of the orbiting body and its shape, e.g., by including quadrupolar terms in the Mathisson-Papapetrou equations of motion.

\appendix

\section{Solving the MP equations: general case}
\label{app1}

The parametric equations for the center of mass line ${\mathcal C}_U$ are given by
\beq
t=t_0+{\hat s}t_{\hat s}\,, \qquad
r=r_0+{\hat s}r_{\hat s}\,, \qquad
\phi=\phi_0+{\hat s}\phi_{\hat s}\,, 
\eeq
so that
\beq
\nu^{\hat r}=\nu^{\hat r}_0+{\hat s}\nu^{\hat r}_{\hat s}\,, \qquad
\nu^{\hat \phi}=\nu^{\hat \phi}_0+{\hat s}\nu^{\hat \phi}_{\hat s}\,, 
\eeq
where all quantities are functions of the proper time $\tau$.
A similar expansion holds for $u$, i.e.,
\beq
\nu_u=\nu_{u0}+{\hat s}\nu_{u{\hat s}}\,, \qquad
\alpha_u=\alpha_{u0}+{\hat s}\alpha_{u{\hat s}}\,, 
\eeq
where $\nu_{u0}=\nu_0$.
The first order correction to $u$ turns out to be 
\begin{eqnarray}\fl
u_{\hat s}&=&\gamma_{u0}\left[\gamma_{u0}\nu_{u0}\nu_{u{\hat s}}n
+(-\nu^{\hat r}_0\alpha_{u{\hat s}}+\cos \alpha_{u0}\nu_{u{\hat s}})e_{\hat r}
+(\nu^{\hat \phi}_0\alpha_{u{\hat s}}+\sin \alpha_{u0}\nu_{u{\hat s}})e_{\hat \phi}\right]\,,\nonumber\\
\fl
&=&\gamma_{u0}{\mathcal P}(U_0)\left[(-\nu^{\hat r}_0\alpha_{u{\hat s}}+\cos \alpha_{u0}\nu_{u{\hat s}})e_{\hat r}
+(\nu^{\hat \phi}_0\alpha_{u{\hat s}}+\sin \alpha_{u0}\nu_{u{\hat s}})e_{\hat \phi}\right]\,,
\end{eqnarray}
where ${\mathcal P}(U_0)$ projects orthogonally to $U_0$.

The zeroth order quantities satisfy Eq.~(\ref{eqPRtest}), i.e.,
\begin{eqnarray}\fl\qquad
\label{eqsordzero}
\frac{\rmd t_0}{\rmd \tau}&=&\frac{\gamma_{u0}}{N_0}\,,\qquad
\frac{\rmd r_0}{\rmd \tau}=\gamma_{u0} N_0 \nu^{\hat r}_0\,,\qquad
\frac{\rmd \phi_0}{\rmd \tau}=\frac{\gamma_{u0}\nu^{\hat \phi}_0}{r_0}\,, \nonumber \\
\fl\qquad
\frac{\rmd \nu_{u0}}{\rmd \tau}
&=& -\zeta_K\nu_K\frac{\sin \alpha_{u0}}{\gamma_{u0}} 
   +\frac{A}{M}\frac{\nu_K^2}{r_0}(1-\nu^{\hat r}_0)(\sin \alpha_{u0} -\nu_{u0})\,,\nonumber \\
\fl\qquad
\frac{\rmd \alpha_{u0}}{\rmd \tau}
&=& \frac{\cos \alpha_{u0}}{\nu_{u0}}\left[\frac{\zeta_K}{\nu_K}\gamma_{u0}(\nu_{u0}^2-\nu_K^2) 
   +\frac{A}{M}\frac{\nu_K^2}{r_0}(1-\nu^{\hat r}_0)\right]\,,
\end{eqnarray}
where 
\beq
\nu^{\hat r}_0=\nu_{u0}\sin\alpha_{u0}\,, \qquad 
\nu^{\hat \phi}_0=\nu_{u0}\cos\alpha_{u0}\,,
\eeq
and the Keplerian value of speed $\nu_K$ and the associated Lorentz factor $\gamma_K$ and angular velocity $\zeta_K$ have been introduced in Eq.~(\ref{nuKdef}).

The first order quantities satisfy the equations 
\begin{eqnarray}\fl\qquad
\label{eqsorduno}
\frac{\rmd t_{\hat s}}{\rmd \tau}&=&\frac{\gamma_{u0}}{N_0}\left[\gamma_{u0}^2(\nu^{\hat r}_0\nu^{\hat r}_{\hat s}+\nu^{\hat \phi}_0\nu^{\hat \phi}_{\hat s})-\frac{\nu_K^2}{r_0}r_{\hat s}\right]\,,\nonumber\\
\fl\qquad
\frac{\rmd r_{\hat s}}{\rmd \tau}&=&\gamma_{u0} N_0 \left\{\gamma_{u0}^2\left[\left(1-(\nu^{\hat \phi}_0)^2\right)\nu^{\hat r}_{\hat s}+\nu^{\hat r}_0\nu^{\hat \phi}_0\nu^{\hat \phi}_{\hat s}\right]+\frac{\nu_K^2}{r_0}\nu^{\hat r}_0r_{\hat s}\right\}\,,\nonumber\\
\fl\qquad
\frac{\rmd \phi_{\hat s}}{\rmd \tau}&=&\frac{\gamma_{u0}}{r_0}\left\{\gamma_{u0}^2\left[\nu^{\hat r}_0\nu^{\hat \phi}_0\nu^{\hat r}_{\hat s}+\left(1-(\nu^{\hat r}_0)^2\right)\nu^{\hat \phi}_{\hat s}\right]-\frac{\nu^{\hat \phi}_0}{r_0}r_{\hat s}\right\}\,,\nonumber\\
\fl\qquad
\frac{\rmd \nu^{\hat r}_{\hat s}}{\rmd \tau}&=&-\left[\left(\frac{N_0^2}{\gamma_K^4}-\nu_K^2\right)\frac{\gamma_{u0}}{r_0^2N_0}\left(1-(\nu^{\hat r}_0)^2\right)-\frac{N_0}{r_0^2}\frac{1}{\gamma_{u0}\gamma_K^2}+A\frac{1+N_0^2}{r_0^3N_0^4}(1-\nu^{\hat r}_0)^2\right]r_{\hat s}\nonumber\\
\fl\qquad
&&+\left[\frac{N_0}{r_0\gamma_K^2}\gamma_{u0}^3\nu^{\hat r}_0\left(1-(\nu^{\hat r}_0)^2\right)-\frac{2N_0^2-1}{r_0N_0}\gamma_{u0}\nu^{\hat r}_0-\frac{2A}{r_0^2N_0^2}(1-\nu^{\hat r}_0)\right]\nu^{\hat r}_{\hat s}\nonumber\\
\fl\qquad
&&+ \frac{N_0\gamma_{u0}\nu^{\hat \phi}_0}{r_0}\left[1+\frac{\gamma_{u0}^2}{\gamma_K^2}\left(1-(\nu^{\hat r}_0)^2\right)\right]\nu^{\hat \phi}_{\hat s}\nonumber\\
\fl\qquad
&&-3M\zeta_K^2\gamma_{u0}\nu^{\hat \phi}_0\left(1-(\nu^{\hat r}_0)^2\right)+\frac{2A^2M}{r_0^4N_0^4}\gamma_{u0}^3\nu^{\hat \phi}_0(1-\nu^{\hat r}_0)^4\nonumber\\
\fl\qquad
&&+\frac{AM}{r_0^3N_0^3}\gamma_{u0}^2\nu^{\hat \phi}_0(1-\nu^{\hat r}_0)\left[\frac{2N_0^2}{\gamma_K^2}\left(1-(\nu^{\hat r}_0)^2\right)+2\nu^{\hat r}_0(1-\nu^{\hat r}_0)-\frac{N_0^2}{\gamma_{u0}^2}\right]\,, \nonumber\\
\fl\qquad
\frac{\rmd \nu^{\hat \phi}_{\hat s}}{\rmd \tau}&=&\left[\left(\frac{N_0^2}{\gamma_K^4}-\nu_K^2\right)\frac{\gamma_{u0}\nu^{\hat r}_0}{r_0^2N_0}+A\frac{1+N_0^2}{r_0^3N_0^4}(1-\nu^{\hat r}_0)\right]\nu^{\hat \phi}_0r_{\hat s}\nonumber\\
\fl\qquad
&&-\left[\frac{N_0}{r_0\gamma_K^2}\gamma_{u0}^3\left(1-(\nu^{\hat \phi}_0)^2\right)-\frac{A}{r_0^2N_0^2}\right]\nu^{\hat \phi}_0\nu^{\hat r}_{\hat s}\nonumber\\
\fl\qquad
&&-\left[\frac{N_0}{r_0\gamma_K^2}\gamma_{u0}^3\nu^{\hat r}_0\left(1-(\nu^{\hat r}_0)^2\right)+\frac{A}{r_0^2N_0^2}(1-\nu^{\hat r}_0)\right]\nu^{\hat \phi}_{\hat s}\nonumber\\
\fl\qquad
&&+3M\zeta_K^2\gamma_{u0}\nu^{\hat r}_0(\nu^{\hat \phi}_0)^2+\frac{2A^2M}{r_0^4N_0^4}\gamma_{u0}^3(1-\nu^{\hat r}_0)^3\left(1-(\nu^{\hat \phi}_0)^2-\nu^{\hat r}_0\right)\nonumber\\
\fl\qquad
&&-\frac{AM}{r_0^3N_0}\left[2\gamma_{u0}^2\nu^{\hat r}_0(1-\nu^{\hat r}_0)^2\left(\frac{1-\nu^{\hat r}_0}{\gamma_K^2}+3\nu^{\hat r}_0\right)\right.\nonumber\\
\fl\qquad
&&\left.-(1-\nu^{\hat r}_0)[1+5\nu^{\hat r}_0-2\nu_K^2(1-\nu^{\hat r}_0)]+\frac{1}{\gamma_{u0}^2}\right]\,.
\end{eqnarray}
The remaining quantities $\nu_{u{\hat s}}$ and $\alpha_{u{\hat s}}$ are related to the first order spatial velocities by the algebraic relations
\begin{eqnarray}\fl\qquad
\label{altre}
\nu_{u{\hat s}}&=&\frac{AM}{r_0^2N_0^2}\frac{\cos \alpha_{u0}}{\gamma_{u0}}(1-\nu^{\hat r}_0)+\sin\alpha_{u0}\nu^{\hat r}_{\hat s}+\cos\alpha_{u0}\nu^{\hat \phi}_{\hat s}\,, \nonumber\\
\fl\qquad
\alpha_{u{\hat s}}&=&\frac{AM}{r_0^2N_0^2}\frac{\gamma_{u0}}{\nu_{u0}}(1-\nu^{\hat r}_0)(\nu_{u0}-\sin \alpha_{u0})+\frac{\cos\alpha_{u0}\nu^{\hat r}_{\hat s}-\sin\alpha_{u0}\nu^{\hat \phi}_{\hat s}}{\nu_{u0}}\,.
\end{eqnarray}

\subsection{Equilibrium solutions}
\label{app1a}

In order to find an equilibrium position at a given point $(r(\tau_\ast),\pi/2,\phi(\tau_\ast))$ for values $\tau\geq\tau_\ast$ of the proper time we have to impose first the conditions $\rmd r/\rmd \tau=0$ and $\rmd \phi/\rmd \tau=0$, which are fulfilled by $\nu^{\hat r}_0=0=\nu^{\hat \phi}_0$ (i.e., $\nu_{u0}=0$) and $\nu^{\hat r}_{\hat s}=0=\nu^{\hat \phi}_{\hat s}$.
Requiring their first derivatives with respect to $\tau$ to be identically vanishing as well gives the condition (\ref{balcond}) from the zeroth order equations (\ref{eqsordzero}).
The first order set (\ref{eqsorduno}) yields
\begin{eqnarray}
\frac{\rmd t_{\hat s}}{\rmd \tau}&=&-\frac{\nu_K^2}{r_0N_0}r_{\hat s}\,,\qquad
\frac{\rmd r_{\hat s}}{\rmd \tau}=0\,,\qquad
\frac{\rmd \phi_{\hat s}}{\rmd \tau}=0\,,\nonumber\\
\phantom{x}0&=&-\zeta_K^2\frac{1+N_0^2}{N_0^4}\left[\frac{A}{M}-N_0^2\left(1-\frac{M}{r_0}\frac{N_0}{1+N_0^2}\right)\right]r_{\hat s}\,, \nonumber\\
\phantom{x}0&=&\frac{2AM}{r_0^4N_0^4}(A-MN_0)\,.
\end{eqnarray}
The last condition is nothing but Eq.~(\ref{balcond}), whereas the previous one implies $r_{\hat s}=0$ at all values of the proper time.
Equilibrium positions are thus characterized by
\beq
U=n\,, \qquad
P=m(n+M{\hat s}\zeta_K\nu_Ke_{\hat \phi})\,,
\eeq
at $r=r_{\rm(crit)}$.

\section{Solving the MP equations: the case of small $f$}
\label{app2}

The quantities first order in ${\hat s}$ satisfy Eqs.~(\ref{eqsorduno}) and (\ref{altre}) where also terms proportional to $f{\hat s}$ are neglected, taking into account that 
\beq
\nu^{\hat r}_0=0\,, \quad  
\nu^{\hat \phi}_0=\pm\nu_K=\nu_{u0}\,, \quad
\gamma_{u0}=\gamma_K\,, \quad
\alpha_{u0}=0\,,
\eeq 
according to Eq.~(\ref{neargeo}).
We have
\begin{eqnarray}
\label{eqshats}
\frac{\rmd t_{\hat s}}{\rmd \tau}&=&\frac{\gamma_K}{N_0}\left[\pm\gamma_K^2\nu_K\nu^{\hat \phi}_{\hat s}-\frac{\nu_K^2}{r_0}r_{\hat s}\right]\,,\nonumber\\
\frac{\rmd r_{\hat s}}{\rmd \tau}&=&\gamma_K N_0\nu^{\hat r}_{\hat s}\,,\nonumber\\
\frac{\rmd \phi_{\hat s}}{\rmd \tau}&=&\frac{\gamma_K}{r_0}\left[\gamma_K^2\nu^{\hat \phi}_{\hat s}\mp\frac{\nu_K}{r_0}r_{\hat s}\right]\,,\nonumber\\
\frac{\rmd \nu^{\hat r}_{\hat s}}{\rmd \tau}&=&\frac{\gamma_K\nu_K^2}{r_0^2N_0}r_{\hat s}\pm2\gamma_K\zeta_K\nu^{\hat \phi}_{\hat s}\mp3M\gamma_K\nu_K\zeta_K^2\,, \nonumber\\
\frac{\rmd \nu^{\hat \phi}_{\hat s}}{\rmd \tau}&=&\mp\frac{\zeta_K}{\gamma_K}\nu^{\hat r}_{\hat s}\,.
\end{eqnarray}
The remaining quantities $\nu_{u{\hat s}}$ and $\alpha_{u{\hat s}}$ are related to the first order spatial velocities by the algebraic relations
\beq
\label{altre2}
\nu_{u{\hat s}}=\nu^{\hat \phi}_{\hat s}\,, \qquad
\alpha_{u{\hat s}}=\pm\frac{\nu^{\hat r}_{\hat s}}{\nu_K}\,.
\eeq

The equations for the quantities first order in $f$ come from the linearization of Eq.~(\ref{eqsordzero}), where also terms proportional to $f{\hat s}$ are neglected
\begin{eqnarray}
\label{eqsf}
\frac{\rmd t_f}{\rmd \tau}&=&\frac{\gamma_K\nu_K}{N_0}\left[\pm\gamma_K^2\nu^{\hat \phi}_f-\frac{\nu_K}{r_0}r_f\right]\,,\nonumber\\
\frac{\rmd r_f}{\rmd \tau}&=&\gamma_K N_0\nu^{\hat r}_f\,,\nonumber\\
\frac{\rmd \phi_f}{\rmd \tau}&=&\frac{\gamma_K}{r_0}\left[\gamma_K^2\nu^{\hat \phi}_f\mp\frac{\nu_K}{r_0}r_f\right]\,,\nonumber\\
\frac{\rmd \nu^{\hat r}_f}{\rmd \tau}&=&\frac{\gamma_K\nu_K^2}{r_0^2N_0}r_f\pm2\gamma_K\zeta_K\nu^{\hat \phi}_f+\frac{\nu_K^2}{r_0}\,, \nonumber\\
\frac{\rmd \nu^{\hat \phi}_f}{\rmd \tau}&=&\mp\left[\frac{\zeta_K}{\gamma_K}\nu^{\hat r}_f+\frac{\nu_K^3}{r_0}\right]\,.
\end{eqnarray}
Finally, the remaining quantities $\nu_{uf}$ and $\alpha_{uf}$ turn out to be simply given by
\beq
\label{altre2f}
\nu_{uf}=\nu^{\hat \phi}_{\hat f}\,, \qquad
\alpha_{uf}=\pm\frac{\nu^{\hat r}_f}{\nu_K}\,.
\eeq

We have now to solve the coupled system of equations (\ref{eqshats}) and (\ref{eqsf}).
Taking the derivative of the equations for the first order radial components of the spatial velocity with respect to proper time yields
\beq
\label{eqnursf}
\frac{\rmd^2 \nu^{\hat r}_{\hat s}}{\rmd \tau^2}+\Omega_{\rm ep}^2\nu^{\hat r}_{\hat s}=0\,, \qquad
\frac{\rmd^2 \nu^{\hat r}_f}{\rmd \tau^2}+\Omega_{\rm ep}^2\nu^{\hat r}_f+2\Gamma_K\nu_K^2\zeta_K^2=0\,,
\eeq
where the epicyclic frequency $\Omega_{\rm(ep)}$ has been introduced in Eq.~(\ref{omegaep}).
The general solution of Eq.~(\ref{eqnursf}) is straightforward
\begin{eqnarray}
\nu^{\hat r}_{\hat s}&=&a_1\sin(\Omega_{\rm ep}\tau)+a_2\cos(\Omega_{\rm ep}\tau)\,, \nonumber\\
\nu^{\hat r}_f&=&b_1\sin(\Omega_{\rm ep}\tau)+b_2\cos(\Omega_{\rm ep}\tau)-2\Gamma_K\nu_K^2\frac{\zeta_K^2}{\Omega_{\rm ep}^2}\,.
\end{eqnarray}
The first order azimuthal components of the spatial velocity are then given by
\begin{eqnarray}\fl\quad
\nu^{\hat \phi}_{\hat s}&=&\pm\frac{\zeta_K}{\gamma_K\Omega_{\rm ep}}[a_1\cos(\Omega_{\rm ep}\tau)-a_2\sin(\Omega_{\rm ep}\tau)]+\frac32M\zeta_K\nu_K\mp\frac{\nu_Kc_1}{2r_0N_0^2}\,, \nonumber\\
\fl\quad
\nu^{\hat \phi}_f&=&\pm\frac{\zeta_K}{\gamma_K\Omega_{\rm ep}}[b_1\cos(\Omega_{\rm ep}\tau)-b_2\sin(\Omega_{\rm ep}\tau)]\mp\frac{\nu_K}{2N_0^2}\left(\frac{\zeta_K}{\Omega_K}+\frac{c_2}{r_0}\right)\pm\frac{\nu_K^3}{r_0}\frac{\Omega_K^2}{\Omega_{\rm ep}^2}\tau\,.
\end{eqnarray}
Finally, the first order corrections to the orbit turn out to be
\begin{eqnarray}\fl\quad
t_{\hat s}&=&2r_0\frac{\Omega_K^2}{\Omega_{\rm ep}^2}[a_1\sin(\Omega_{\rm ep}\tau)+a_2\cos(\Omega_{\rm ep}\tau)]-\frac32\frac{\gamma_K^3\nu_K\zeta_K}{N_0^2}[c_1\mp MN_0\nu_K]\tau+d_1\,, \nonumber\\
\fl\quad
t_f&=&2r_0\frac{\Omega_K^2}{\Omega_{\rm ep}^2}[b_1\sin(\Omega_{\rm ep}\tau)+b_2\cos(\Omega_{\rm ep}\tau)]-\frac{\gamma_K^2}{2r_0N_0^4}[M+3\gamma_K\nu_K^2N_0^3c_2]\tau\nonumber\\
\fl\quad
&&+\frac32M\Gamma_K\frac{\Omega_K^4}{\Omega_{\rm ep}^2}\tau^2+d_2\,, \nonumber\\
\fl\quad
r_{\hat s}&=&\frac{N_0\gamma_K}{\Omega_{\rm ep}}[a_2\sin(\Omega_{\rm ep}\tau)-a_1\cos(\Omega_{\rm ep}\tau)]+c_1\,, 
\nonumber\\
\fl\quad
r_f&=&\frac{N_0\gamma_K}{\Omega_{\rm ep}}[b_2\sin(\Omega_{\rm ep}\tau)-b_1\cos(\Omega_{\rm ep}\tau)]-2\frac{M}{r_0}\frac{\Omega_K^2}{\Omega_{\rm ep}^2}\tau+c_2\,, \nonumber\\
\fl\quad
\phi_{\hat s}&=&\pm\frac{2\gamma_K^2\zeta_K}{r_0\Omega_{\rm ep}^2}[a_1\sin(\Omega_{\rm ep}\tau)+a_2\cos(\Omega_{\rm ep}\tau)]\mp\frac32\frac{\gamma_K^3\zeta_K^2}{N_0^2\nu_K}[c_1\mp MN_0\nu_K]\tau+e_1\,, 
\nonumber\\
\fl\quad
\phi_f&=&\pm\frac{2\gamma_K^2\zeta_K}{r_0\Omega_{\rm ep}^2}[b_1\sin(\Omega_{\rm ep}\tau)+b_2\cos(\Omega_{\rm ep}\tau)]\mp\frac{\gamma_K^2}{2r_0^2N_0^3}[M+3\gamma_K\nu_K^2N_0^3c_2]\tau\nonumber\\
\fl\quad
&&\pm\frac32r_0\zeta_K^2\frac{\Omega_K^3}{\Omega_{\rm ep}^2}\tau^2+e_2\,,
\end{eqnarray}
where $a_1$, $a_2$, $b_1$, $b_2$, $c_1$, $c_2$, $d_1$, $d_2$, $e_1$, $e_2$ are arbitrary integration constants.
Their values are fixed by requiring that all first order quantities vanish at $\tau=0$.
The corresponding solution is given by Eqs.~(\ref{solfs1}) and  (\ref{solfs2}).

\section*{Acknowledgement}
The authors are indebted to Profs. R.T. Jantzen and L. Stella for initiating the discussion of the Poynting-Robertson effect in general relativity. 
ICRANet is thanked for support.

\end{document}